\documentclass[apjl]{emulateapj}
\usepackage[colorlinks,urlcolor=blue,citecolor=blue,linkcolor=blue]{hyperref}

\def \solar{_\odot}
\def \pow10#1{\times 10^{#1}}

\tighten

\begin{document}

\submitted{ApJ Letters accepted version}

\title{Observations of dark and luminous matter: the radial
  distribution of satellite galaxies around massive red galaxies}

\author{Tomer Tal\altaffilmark{1},
  David A. Wake\altaffilmark{1},
  Pieter G. van Dokkum\altaffilmark{1}}
\altaffiltext{1}{Yale University Astronomy Department, P.O. Box
  208101, New Haven, CT 06520-8101 USA}

\shorttitle{Radial number density profile of LRG satellites}
\shortauthors{Tal, Wake \& van Dokkum}

\begin{abstract}
  We study the projected radial distribution of satellite galaxies
  around more than 28,000 Luminous Red Galaxies (LRGs) at
  $0.28<z<0.40$ and trace the gravitational potential of LRG groups in
  the range $15<r/$kpc$<700$.
  We show that at large radii the satellite number density profile is
  well fitted by a projected NFW profile with $r_s\sim270$ kpc
  and that at small radii this model underestimates the number of
  satellite galaxies.
  Utilizing the previously measured stellar light distribution of LRGs
  from deep imaging stacks we demonstrate that this small scale excess
  is consistent with a non-negligible baryonic mass contribution to
  the gravitational potential of massive groups and clusters.
  The combined NFW+scaled stellar profile provides an excellent fit
  to the satellite number density profile all the way from 15 kpc to
  700 kpc.
  Dark matter dominates the total mass profile of LRG halos at $r>25$
  kpc whereas baryons account for more than 50\% of the mass at
  smaller radii.
  We calculate the total dark-to-baryonic mass ratio and show
  that it is consistent with measurements from weak lensing for
  environments dominated by massive early type galaxies.
  Finally, we divide the satellite galaxies in our sample into three
  luminosity bins and show that the satellite light profiles of all
  brightness levels are consistent with each other outside of roughly
  25 kpc.
  At smaller radii we find evidence for a mild mass segregation with an
  increasing fraction of bright satellites close to the central LRG.
\end{abstract}


\section{\label{intro}Introduction}
 Measurements of the mass profile of galaxy groups and clusters hold
 key insight into the distribution of dark and baryonic matter in
 galaxy environments.
 Such analyses are difficult to perform observationally as the
 gravitational potential of massive environments is predicted to be
 dominated by dark matter halos.
 Therefore, significant efforts have been devoted to indirect
 measurements, such as observations of the hot interstellar medium and
 the dynamics of satellite galaxies in groups and clusters
 \citep[e.g.,][]{fabricant_x-ray_1980, forman_x-ray-imaging_1982,
 mulchaey_diffuse_1993, girardi_velocity_1993,
 fadda_observational_1996, vikhlinin_chandra_2006}.
 Other indirect approached, such as galaxy clustering and
 gravitational lensing, successfully measure the distribution of mass
 through its effect on other mass concentrations
 \citep[e.g.,][]{vader_small-scale_1991, zehavi_galaxy_2002,
   madore_companions_2004, mandelbaum_galaxy_2006, 
   masjedi_very_2006, gavazzi_sloan_2007, masjedi_growth_2008,
   bolton_sloan_2008, wake_2df-sdss_2008, watson_extreme_2011}.
 
 One such method utilizes measurements of the radial number density
 profile of satellite galaxies to trace the underlying gravitational
 potential of group or cluster halos.
 This approach is challenging as it requires identifying galaxies as
 satellites and separating them from other objects along the light of
 sight.
 Accurate identifications of satellites using spectroscopic redshifts
 are observationally expensive and are restricted to the brightest
 galaxies.

 Alternatively, one can measure the radial number density profile of 
 satellite galaxies in a statistical manner, in effect stacking
 results from a large number of groups and clusters in a well defined
 sample.
 This method can provide a measurement of the total mass profile at
 essentially all relevant scales but requires a large-scale
 homogeneous imaging samples such as that provided by the Sloan 
 Digital Sky Survey \citep[SDSS;][]{york_sloan_2000}.

 Throughout the paper we adopt the following cosmological parameters:
 $\Omega_m=0.3$, $\Omega_{\Lambda}=0.7$ and $H_0=70$ km s$^{-1}$
 Mpc$^{-1}$.

 \begin{figure*}
   \center
   \includegraphics[width=1.0\textwidth]{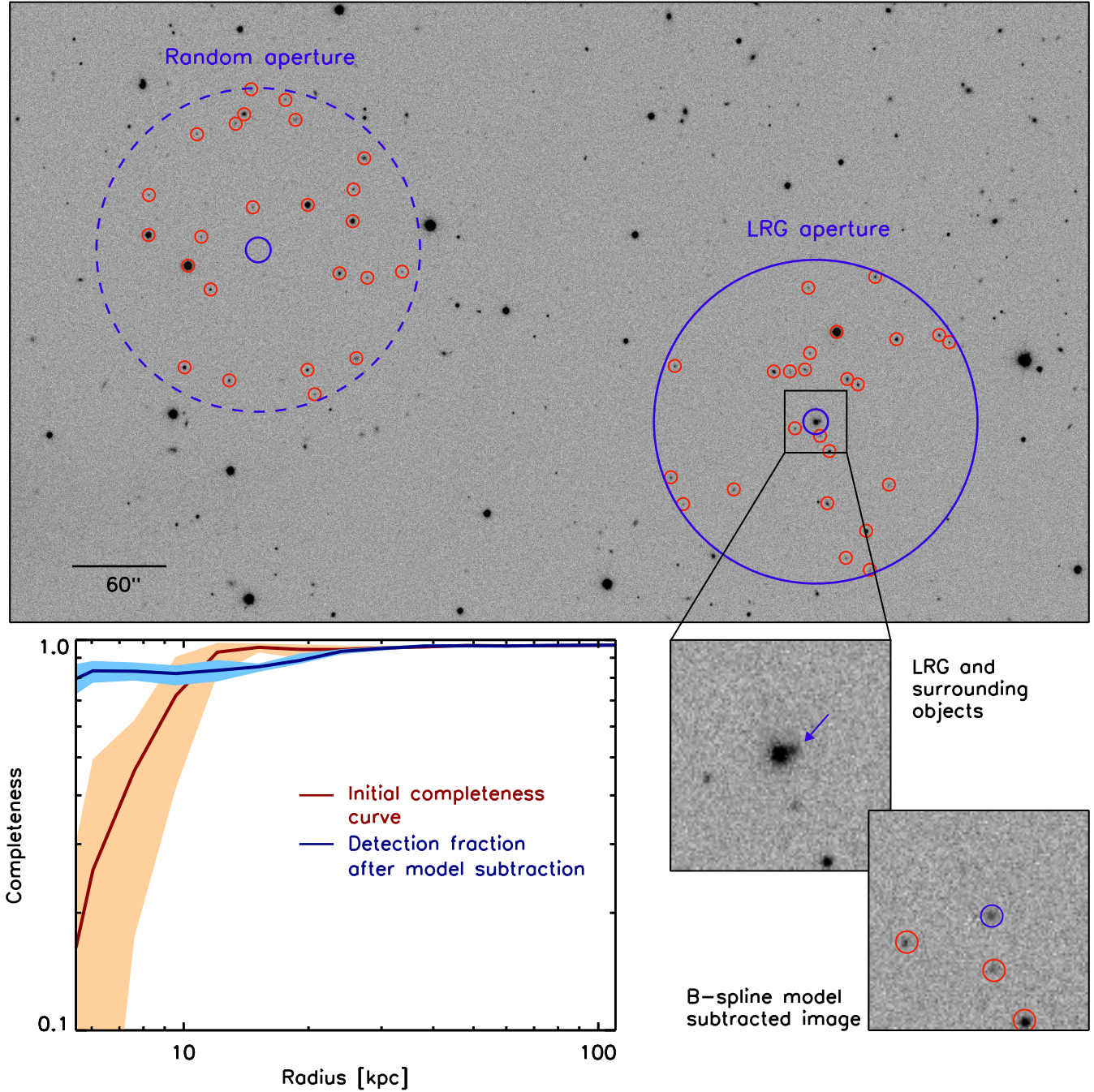}
   \caption{Top panel: illustration of the satellite radial profile
     extraction technique.
     Objects are identified in annuli around the LRG (red circles) and
     projected distances are measured from the central galaxy.
     An identical analysis is then performed in randomly positioned
     apertures along the same SDSS fields and object distances are
     measured to the center of the random aperture.
     After combining the measurements of more than 28,000 apertures we
     subtracted the contribution of foreground and background sources
     from the radial profile of objects in LRG environments.
     Bottom right: upon subtraction of a b-spline model fit to the
     central LRG, additional objects may be detected at small
     projected radii (blue arrow and blue circle).
     Bottom left: completeness tests using mock galaxies show that
     model fitting and subtraction uncovers a large fraction of
     satellite galaxies at small projected radii.
     The blue and orange curves represent the distribution of
     completeness levels for the full satellite brightness range for
     galaxies modeled with a S\'{e}rsic profile with $n=$ 2.5 and
     $r_e=$ 4 kpc.
   }
   \label{fig:examp}
 \end{figure*}

\section{Data and analysis}
 \label{sec:data}
 We selected galaxy images for this study from the 7th data release of
 SDSS \citep{abazajian_seventh_2009} using the criteria presented in
 \cite{tal_mass_2012}.
 This sample includes the reddest, most luminous LRGs at redshift
 $0.28<z<0.4$ with restframe $g-$band magnitudes brighter than -22.42
 and restframe $g-r$ color values redder than 0.74.
 These are the most massive galaxies in the nearby universe with
 total stellar masses greater than a few times $10^{11}$M$_{\solar}$.
 More than 90\% of the selected LRGs are expected to be the central
 galaxies in their halos, and are thought to be residing in groups
 with typical total masses of a few times $10^{13}$M$_{\solar}$ 
 \citep{wake_2df-sdss_2008, zheng_halo_2009, reid_constraining_2009}.
 The final data set includes 28,017 LRGs for which we acquired
 $r-$band images from the SDSS archive.

 In order to extract the radial distribution of satellite galaxies
 in LRG environments we followed a statistical approach to remove
 the contamination of foreground and background objects from the LRG
 fields.
 We started by subtracting the LRG light from each image using a
 fourth-order radial b-spline model, utilizing the IDLUTIL package
 \citep{bolton_sloan_2006}.
 Next, we used SExtractor \citep{bertin_sextractor:_1996} to detect
 all objects in an annulus of radius 150'' (720 kpc at $z=0.34$)
 around the LRGs, utilizing a detection threshold of 2$\sigma$ above
 the background.
 We measured the projected physical distance between each detected
 object and the central LRG, assuming that all sources are at the same
 redshift as the LRG.
 We then combined the results from all fields and divided them
 into radial bins.
 Finally, we performed an identical source extraction in 10 randomly 
 positioned apertures along each of the SDSS fields and measured
 projected distances to the aperture centers.
 An illustration of the measurement technique is presented in
 Figure \ref{fig:examp}.

 The number density of objects detected in the proximity of any
 bright source typically underestimates the true number of
 sources in that region.
 To quantify the success rate of object detection close to
 the LRGs we randomly positioned five mock galaxy images in each SDSS
 frame and repeated the analysis portrayed in Section \ref{sec:data}.
 We modeled these galaxies using a S\'{e}rsic light profile with
 index $n=2.5$, effective radius $r_e=4$ kpc and total brightness
 between 18.7 and 21.3 magnitudes.
 The bottom-left panel of Figure \ref{fig:examp} shows the detection
 fraction of mock galaxies from the LRG fields before and after
 subtracting the best-fitting b-spline model.
 We derived a correction factor from the running median of the model
 subtracted curve (blue line in Figure \ref{fig:examp}) to account
 for missing objects due to incompleteness.
 In addition to demonstrating a high completeness rate of more than
 80\% at projected $r>6$ kpc, this figure shows the efficiency of 
 b-spline model fitting at characterizing the central galaxy light at
 all radii \citep[in agreement with results by][]{bolton_sloan_2006,
   nierenberg_luminous_2011}.\\
 \\
 
 \begin{figure}
   \includegraphics[width=0.47\textwidth]{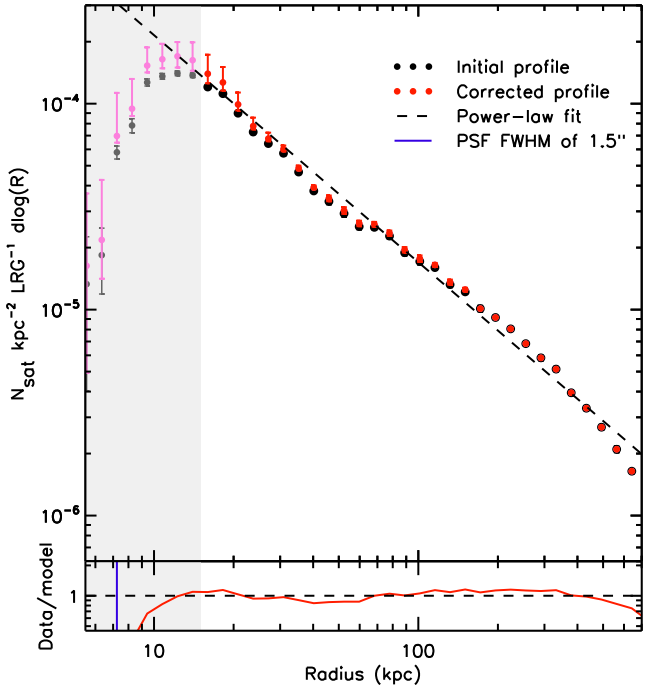}
   \caption{Projected number density profile of satellite galaxies
     around LRGs at $z=0.34$.
     The black points show the radial distribution of sources after
     a statistical subtraction of foreground and background objects.
     The red points represent the same distribution after a correction
     factor was applied to account for incompleteness.
     At $15<r/$kpc$<700$ the profile is reasonably well fitted by a
     single power-law model with a slope of -1.1 and residuals of up
     to roughly 20\% (dashed line).
     Data points inside of the fitting radius (15 kpc) are marked by
     a shaded region and the full width half maximum of typical SDSS
     stellar point spread function is represented by a blue line.
   }
   \hfill
   \label{fig:fig2}
 \end{figure}

\section{Results}
\label{sec:results}
 The resulting projected number density profile of satellite galaxies
 around LRGs is shown in Figure \ref{fig:fig2}.
 Black points represent the ``raw'' difference between counts in
 LRG fields and in random fields and red points are corrected for
 incompleteness (see Section \ref{sec:data}).
 We show the profile for all sources that are brighter than 21.3
 magnitudes, which is roughly 3 magnitudes fainter than the 
 mean LRG brightness.
 
 In order to assess the errors in the measured profile we
 calculated the standard deviation of the distribution of random
 aperture measurements in each radial bin.
 In addition, we repeated the completeness calculation that is
 described in Section \ref{sec:data} for galaxies modeled with
 S\'{e}rsic parameters between $1<n<4$ and between $2<r_e/$kpc$<6$,
 with total brightness between 18.7 and 21.3 magnitudes.
 The range of completeness correction factor values represents
 a potential systematic error which is driven by the possible spread
 of satellite galaxy properties around the studied LRGs.
 The error bars in Figure \ref{fig:fig2} fold in the uncertainties
 from both of the above mentioned potential error sources.

 The projected number density profile of LRG satellite galaxies is
 confidently traced in the range $15<r/$kpc$<700$.
 At radii larger than roughly 15 kpc the profile declines
 monotonically while at smaller radii the slope is positive.
 Figure \ref{fig:fig2} shows that in the range $15<r/$kpc$<700$ the
 profile is relatively well fitted by a single power-law model of the
 form $\log{N_{sat}}=-1.1\log{r}+2.7\pow10{-3}$ (dashed line) with
 residuals of up to roughly 20\% (bottom panel).
 The gray line in Figure \ref{fig:fig2} shows the full width half
 maximum of the typical stellar point spread function.
 
 \begin{figure*}
   \center
   \includegraphics[width=0.87\textwidth]{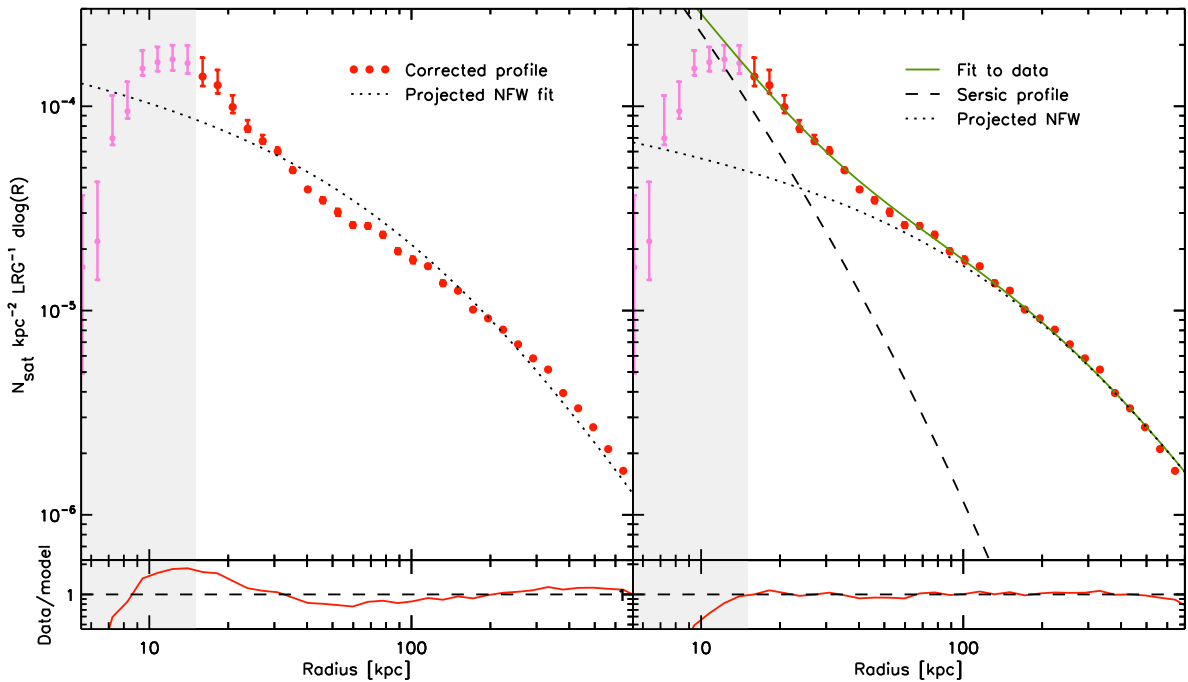}
   \caption{Model fitting to the projected number density profile of
     satellite galaxies around LRGs.
     Left panel: the overall profile (red data points) is poorly
     fitted by a single projected NFW model (black dotted line) with
     residuals of up to roughly 100\%.
     Right panel: an NFW+S\'{e}rsic model fit to the profile in the
     range $15<r/$kpc$<700$.
     At large radii the profile is well fitted by a single
     projected-NFW model (black dotted line);
     when combined with an arbitrarily normalized S\'{e}rsic model
     (black dashed line; $n$ and $r_e$ taken from a fit to the light
     profile of deep imaging stack of the same LRGs), the fit to the
     number density profile is excellent (green solid line).
     The shaded region marks the minimum fitting radius in both
     panels.
   }
   \hfill
   \label{fig:fig3}
 \end{figure*}
 
\section{Model fitting}
 The power-law fit presented in Figure \ref{fig:fig2} provides a
 reasonable description of the projected number-density profile of LRG
 satellites.
 Nevertheless, the residuals from this fit (bottom panel of Figure
 \ref{fig:fig2}) suggest that this model may not best represent the
 projected total mass distribution in LRG environments.
 A more physically motivated model for this may be the NFW profile
 \citep{navarro_structure_1996, navarro_universal_1997}, which
 provides a good approximation of the radial density distribution of
 dark matter halos.
 In the following subsections we test how well the NFW model fits the
 satellite galaxy number density profile.

 \subsection{NFW profile fit}
  Following \cite{bartelmann_arcs_1996} we write the analytical
  approximation of the projected NFW profile as
  \begin{equation}
    \Sigma(x) \propto \left\{ \begin{array}{lc}
      (x^2-1)^{-1}\left(1-\frac{2}{\sqrt{x^2-1}} \mbox{arctan}
      \sqrt{\frac{x-1}{x+1}}\right) & 
      (x>1) \\
      1/3 & (x=1)\\
      (x^2-1)^{-1}\left(1-\frac{2}{\sqrt{1-x^2}} \mbox{arctanh}
      \sqrt{\frac{1-x}{1+x}}\right) & (x<1)
    \end{array}\right.
  \end{equation}
  where $x\equiv r/r_s$ and $r_s$ is the scale radius.
  We used the non-linear least squares curve fitting program MPFIT
  \citep{markwardt_non-linear_2009} to fit the above approximation of
  the projected NFW model to the satellite number-density profile.  
  We allowed both model parameters ($r_s$ and the normalization
  factor) to be fitted, in practice allowing the fit to be arbitrarily
  normalized.
  The left panel of Figure \ref{fig:fig3} shows that a single NFW
  profile (black dotted line) fails to describe the full range of the
  number-density profile (red points) better than the power-law
  model.
  This is especially relevant on small scales ($r<25$ kpc), where the
  projected NFW model underpredicts the number of satellite galaxies
  by up to a factor of two (bottom left panel).
  Assuming that the number density profile traces the total mass in
  LRG halos, this may suggest that an additional component contributes
  to the gravitational potential on small scales.

 \subsection{NFW+S\'{e}rsic}
  Although baryons are predicted to contribute only a small
  fraction of the total mass of massive halos, they are more
  concentrated than the dark matter halos with stellar
  half-light radii of roughly 10 kpc.
  This implies that on small scales the gravitational potential of LRG
  groups and clusters may be dominated by baryons, whose distribution
  does not necessarily follow an NFW profile.
  
  To test this we utilized the surface brightness profile of a deep
  imaging stack of more than 40,000 $z=$0.34 LRGs from
  \cite{tal_faint_2011}.
  The stacked light profile portrays the distribution of luminous
  matter in the galaxy and it is well fitted by a single S\'{e}rsic
  model in the range $1<r/$kpc$<100$.
  We utilized the best-fit S\'{e}rsic parameters from the stacking
  analysis ($n=5.5$ and $r_e=13.1$ kpc) to fit the satellite
  number-density profile with an NFW+S\'{e}rsic function.
  This fit has three free parameters: the NFW scale radius $r_s$ and
  independent normalizations for each of the two model components.

  The upper right panel of Figure \ref{fig:fig3} shows the best fit
  NFW+S\'{e}rsic model (green line) to the number-density profile (red
  points).
  The bottom right panel of the figure shows that this model provides
  an excellent description of the projected number density profile
  with fit residuals of less than 10\%.
  At large radii the profile is well fitted by an NFW model with
  $r_s=267\pm6$ kpc (black dotted line) while at small radii the
  excess number of satellites is dominated by the S\'{e}rsic model
  (black dashed line).
  The error bars in Figure \ref{fig:fig3} fold in the same
  uncertainties as those discussed in Section \ref{sec:results} and
  they show that the shape of the satellite number density profile
  depends only mildly on the satellite galaxy properties.
  
  Finally, we note that for a typical LRG group halo mass of
  $5\pow10{13}$ M$\solar$ the best-fit scale radius implies a
  concentration parameter $c\equiv r_{200}/r_s\sim2$.
  The quoted uncertainty in the value of $r_s$ is simply the formal
  1$\sigma$ fitting error and does not account for systematic errors.
  We discuss the effect of such systematic uncertainties in Section
  \ref{sec:errors}.

  \begin{figure}
   \includegraphics[width=0.47\textwidth]{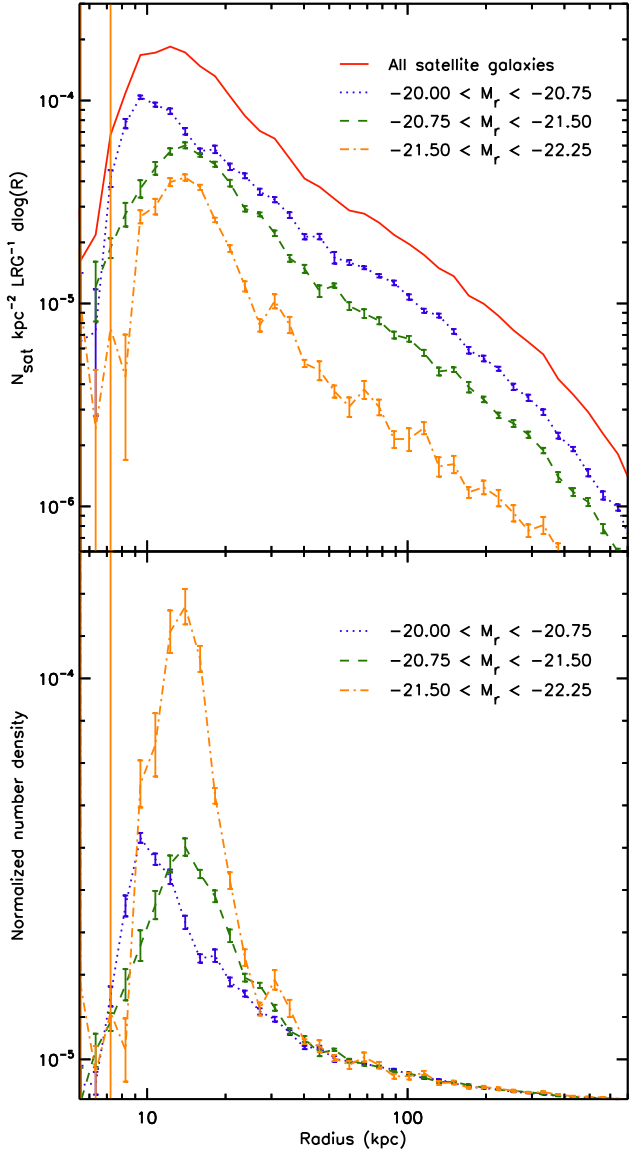}
   \caption{Projected number density profiles of LRG satellites binned
     by luminosity.
     Top panel: the solid red line shows the overall profile of all
     satellite galaxies.
     Dotted purple, dashed green and dot-dashed orange lines show the
     radial satellite distribution of the faintest, intermediate
     luminosity and most luminous galaxies, respectively.
     Bottom panel: The three profiles are normalized to the
     intermediate luminosity bin according to their $r\sim200$ kpc
     data point.
     While the three curves agree in the range $25<r/$kpc$<700$, the
     fraction of luminous satellites increases significantly at small
     projected radii.
     All lines are plotted in log-linear scale to enhance the
     difference between the profiles.
   }
   \hfill
   \label{fig:fig4}
 \end{figure}
 
\section{Discussion}
\label{sec:discussion}
 \subsection{Ratio of dark to luminous matter}
  The excellent agreement between the projected number density profile
  of satellite galaxies around LRGs and an NFW+S\'{e}rsic model
  suggests that the contribution of baryonic mass is non-negligible in
  the inner regions of such groups and clusters.
  Assuming that the satellite profile traces the total mass 
  distribution we can utilize the functional fits to estimate the ratio
  between dark and baryonic mass in LRG environments.
  To do so we separately integrated the two components of the best-fit
  NFW+S\'{e}rsic model out to 700 kpc and calculated the area under
  the curves, resulting in a total dark-to-baryonic mass ratio of 80.
  If we further assume that stars typically account for only half of
  the baryonic mass in galaxy groups \citep{mulchaey_x-ray_2000} then
  the implied baryon conversion factor, $\eta=0.037$, is consistent
  with the weak lensing result from \cite{mandelbaum_galaxy_2006} for
  massive early type galaxies.

  The high fraction of baryonic mass in LRG groups may suggest that
  LRGs represent an extreme population of galaxies, where a large
  fraction of the stellar mass in the halo had been incorporated
  into the central galaxy.
  This is in agreement with \cite{tal_mass_2012} who found that roughly
  35\% of the stellar mass in LRG groups in locked in the LRG itself.
  Recent numerical and observational studies of satellites around
  central galaxies with a broader range of properties find a less
  pronounced deviation from a projected NFW profile
  \citep[][]{van_den_bosch_abundance_2005, sales_satellite_2007,
  guo_satellite_2012}.
 
 \subsection{Small scale number density profile}
  Throughout the analysis of the number density profile we utilized a
  lower radius limit for the functional fits in order to avoid the
  inner part of the profile at $r<15$ kpc.
  At small radii the number density of satellite galaxies drops
  dramatically despite a high expected detection fraction (as
  predicted by the completeness curve of Figure \ref{fig:examp}).
  We consider two plausible scenarios to explain this change in
  profile slope: a deficit of satellite galaxies due to interactions
  with the LRG and an underestimate of the true number of satellites due
  to detection limits.

  The turnover point of the profile slope is at roughly 15kpc, only
  about 2 kpc farther out than the average half-light radius of LRGs
  \citep[13.1 kpc;][]{tal_mass_2012}.
  In such proximity to the massive central, satellite galaxies
  experience strong tidal forces and possibly ram-pressure stripping,
  both acting to remove mass from the infalling galaxies.
  It is possible that many satellite galaxies below a certain
  mass threshold do not survive for long close to the halo center and
  that their stars are quickly incorporated into the LRG itself.

  Alternatively, our simulations may underestimate the effects of
  incompleteness close to the LRG itself.
  However, we note that the deficiency of satellites at small
  radii cannot be caused by simple signal-to-noise ratio effects as
  this deficiency seems to be strongest for the brightest galaxies
  (see below).
  To test this, imaging of higher spatial resolution and a more
  accurate modeling of individual LRGs are needed to resolve the
  satellite profile on the smallest scales.
  Such analyses would help determine whether the drop we find in the
  profile at $r<15$ kpc is indeed real.

 \subsection{Dependence on satellite mass}
 \label{sec:mass_seg}
  The projected number-density profile can be further utilized to
  study the distribution of mass in LRG halos.
  We divided our sample of detected objects into three luminosity bins
  (assuming all objects are at the same redshift as the LRG in their
  field) and repeated the statistical number density profile
  extraction for each bin.
  Figure \ref{fig:fig4} shows the resulting profiles of the most
  luminous, intermediate luminosity and faintest satellite galaxies
  (orange, green and blue lines, respectively).
  In the top panel the relative contributions of each luminosity bin
  is presented, as well as the overall profile of all satellites
  (solid red line).
  In the (linearly rescaled) bottom panel we normalized the three
  profiles to the intermediate luminosity bin based on the curve value
  at $r\sim200$ kpc.
  
  Figure \ref{fig:fig4} shows that at $r>25$ kpc the normalized
  profiles agree remarkably well with each other while at smaller
  radii the fraction of bright satellite galaxies increases.
  The good agreement between the three luminosity bins in the range
  $25<r/$kpc$<700$ suggests that there does not exist a strong mass
  segregation of LRG satellites.
  We also note that in this range the overall number density profile
  is dominated by the dark matter halo (Figure \ref{fig:fig3}).
  At radii smaller than 25 kpc, where baryons dominate the
  gravitational potential, the fraction of luminous satellites
  seems to increase.
  A na\"{i}ve interpretation would suggest that as satellite galaxies
  get closer to the central LRG the probability of retaining their
  stars is proportional to their initial mass.
  However, since the difference between the three luminosity bins is
  only pronounced in the innermost parts of the profile, we conclude
  that our results are consistent with only a mild mass segregation in
  LRG environments.

 \subsection{Sources of uncertainty}
 \label{sec:errors}
  Throughout this paper we utilize a key assumption that satellite
  galaxies around LRGs are an unbiased tracer of the total mass
  distribution.  
  However, it is known that halos, especially at the high-mass end of
  the mass function, continue to evolve and are not in dynamical
  equilibrium \citep[e.g.,][]{bell_dry_2006, wake_2df-sdss_2008,
  tal_frequency_2009, van_dokkum_growth_2010,
  tojeiro_disentangling_2011, skelton_modeling_2011}.
  This implies that measurements of projected satellite radii (and 
  therefore the number density profile) may potentially be dominated
  by, e.g., the trajectories of infalling systems.
  Estimates of the dynamical state of LRG halos are important
  for proper assessment of our results but are unfortunately beyond
  the scope of this paper.

  Furthermore, we note that our measurement of background and
  foreground sources was not extracted from truly random fields.
  The random apertures which we used to derive this measurement were
  positioned along the same SDSS fields as the LRG apertures to
  match the photometric properties of the fields and to sample the
  large-scale structure around the LRG halos.
  This implies that some satellite galaxies were possibly included in
  the random aperture catalogs, resulting in an over-subtraction of 
  background and foreground sources, especially at large radii.

  An additional source of uncertainty stems from our selection of
  the model fitting range.
  In order to test how the best-fit scale radius of the NFW
  profile depends on the choice of fitting boundaries we varied the
  lower threshold radius between 10 and 25 kpc and repeated the
  NFW+S\'{e}rsic model fitting procedure.
  The resulting scale radius was found to be in the range
  $240<r_s/kpc/270$ and the stellar conversion efficiency remained
  in the range $0.030<\eta<0.037$.
  The small variation in derived $r_s$ and $\eta$ suggests that the
  number density profile is only mildly sensitive to the choice of
  lower fitting limit.

  The magnitude threshold used for source extraction also affects the
  resulting best-fit model parameters.
  To test this we extended our analysis to fainter satellites and
  found that the dark-to-baryonic mass ratio increases by up to 20\%
  when the extraction limit is set to 22 magnitudes.
  However, we note that the detection fraction of sources in this
  brightness range drops to under 70\% at large radii and to roughly
  50\% at $r=15$ kpc.
  This suggests that the above mentioned increase in the value of
  dark-to-baryonic mass ratio is likely dominated by underestimates of
  the number density of faint satellites.
  
  Finally, the physical processes that affect satellite galaxies at
  small radii (such as tidal stripping) may also influence galaxies,
  to a lesser degree, everywhere else in the group or cluster.
  This means that satellites may become fainter as they gradually move
  closer to the central LRG.
  Such satellites are more likely to fall under the detection
  threshold with decreasing radius, thus making the number density
  slope overall shallower.

\begin{acknowledgements}
  We gratefully acknowledge support from the CT Space Grant.

  Funding for the SDSS and SDSS-II has been provided by the Alfred
  P. Sloan Foundation, the Participating Institutions, the National
  Science Foundation, the U.S. Department of Energy, the National
  Aeronautics and Space Administration, the Japanese Monbukagakusho,
  the Max Planck Society, and the Higher Education Funding Council for
  England.

\end{acknowledgements}
 
\bibliography{ms}

\begin{thebibliography}{38}
\expandafter\ifx\csname natexlab\endcsname\relax\def\natexlab#1{#1}\fi

\bibitem[{Abazajian {et~al.}(2009)Abazajian, Allende~Prieto, An, Anderson,
  Anderson, Annis, Bahcall, Cunha, Grebel, Okamura, Jorgensen, Zehavi, \&
  York}]{abazajian_seventh_2009}
Abazajian, K.~N., Allende~Prieto, C., An, D., {et~al.} 2009,
  \href{http://adsabs.harvard.edu/abs/2009ApJS..182..543A}{The Astrophysical
  Journal Supplement Series, 182, 543}

\bibitem[{Bartelmann(1996)}]{bartelmann_arcs_1996}
Bartelmann, M. 1996,
  \href{http://adsabs.harvard.edu/abs/1996A%26A...313..697B}{Astronomy and
  Astrophysics, 313, 697}

\bibitem[{Bell {et~al.}(2006)Bell, Naab, {McIntosh}, Somerville, Caldwell,
  Barden, Wolf, Rix, Beckwith, Borch, Häussler, Heymans, Jahnke, Jogee,
  Koposov, Meisenheimer, Peng, Sanchez, \& Wisotzki}]{bell_dry_2006}
Bell, E.~F., Naab, T., {McIntosh}, D.~H., {et~al.} 2006,
  \href{http://adsabs.harvard.edu/abs/2006ApJ...640..241B}{The Astrophysical
  Journal, 640, 241}

\bibitem[{Bertin \& Arnouts(1996)}]{bertin_sextractor:_1996}
Bertin, E., \& Arnouts, S. 1996,
  \href{http://adsabs.harvard.edu/abs/1996A%26AS..117..393B}{Astronomy and
  Astrophysics Supplement Series, 117, 393}

\bibitem[{Bolton {et~al.}(2006)Bolton, Burles, Koopmans, Treu, \&
  Moustakas}]{bolton_sloan_2006}
Bolton, A.~S., Burles, S., Koopmans, L. V.~E., Treu, T., \& Moustakas, L.~A.
  2006, \href{http://adsabs.harvard.edu/abs/2006ApJ...638..703B}{The
  Astrophysical Journal, 638, 703}

\bibitem[{Bolton {et~al.}(2008)Bolton, Treu, Koopmans, Gavazzi, Moustakas,
  Burles, Schlegel, \& Wayth}]{bolton_sloan_2008}
Bolton, A.~S., Treu, T., Koopmans, L. V.~E., {et~al.} 2008,
  \href{http://adsabs.harvard.edu/abs/2008ApJ...684..248B}{The Astrophysical
  Journal, 684, 248}

\bibitem[{Fabricant {et~al.}(1980)Fabricant, Lecar, \&
  Gorenstein}]{fabricant_x-ray_1980}
Fabricant, D., Lecar, M., \& Gorenstein, P. 1980,
  \href{http://adsabs.harvard.edu/abs/1980ApJ...241..552F}{The Astrophysical
  Journal, 241, 552}

\bibitem[{Fadda {et~al.}(1996)Fadda, Girardi, Giuricin, Mardirossian, \&
  Mezzetti}]{fadda_observational_1996}
Fadda, D., Girardi, M., Giuricin, G., Mardirossian, F., \& Mezzetti, M. 1996,
  \href{http://adsabs.harvard.edu/abs/1996ApJ...473..670F}{The Astrophysical
  Journal, 473, 670}

\bibitem[{Forman \& Jones(1982)}]{forman_x-ray-imaging_1982}
Forman, W., \& Jones, C. 1982,
  \href{http://adsabs.harvard.edu/abs/1982ARA%26A..20..547F}{Annual Review of
  Astronomy and Astrophysics, 20, 547}

\bibitem[{Gavazzi {et~al.}(2007)Gavazzi, Treu, Rhodes, Koopmans, Bolton,
  Burles, Massey, \& Moustakas}]{gavazzi_sloan_2007}
Gavazzi, R., Treu, T., Rhodes, J.~D., {et~al.} 2007,
  \href{http://adsabs.harvard.edu/abs/2007ApJ...667..176G}{The Astrophysical
  Journal, 667, 176}

\bibitem[{Girardi {et~al.}(1993)Girardi, Biviano, Giuricin, Mardirossian, \&
  Mezzetti}]{girardi_velocity_1993}
Girardi, M., Biviano, A., Giuricin, G., Mardirossian, F., \& Mezzetti, M. 1993,
  \href{http://adsabs.harvard.edu/abs/1993ApJ...404...38G}{The Astrophysical
  Journal, 404, 38}

\bibitem[{Guo {et~al.}(2012)Guo, Cole, Eke, \& Frenk}]{guo_satellite_2012}
Guo, Q., Cole, S., Eke, V., \& Frenk, C. 2012, Satellite Galaxy Number Density
  Profiles in the Sloan Digital Sky Survey

\bibitem[{Madore {et~al.}(2004)Madore, Freedman, \&
  Bothun}]{madore_companions_2004}
Madore, B.~F., Freedman, W.~L., \& Bothun, G.~D. 2004,
  \href{http://adsabs.harvard.edu/abs/2004ApJ...607..810M}{The Astrophysical
  Journal, 607, 810}

\bibitem[{Mandelbaum {et~al.}(2006)Mandelbaum, Seljak, Kauffmann, Hirata, \&
  Brinkmann}]{mandelbaum_galaxy_2006}
Mandelbaum, R., Seljak, U., Kauffmann, G., Hirata, C.~M., \& Brinkmann, J.
  2006, \href{http://adsabs.harvard.edu/abs/2006MNRAS.368..715M}{Monthly
  Notices of the Royal Astronomical Society, 368, 715}

\bibitem[{Markwardt(2009)}]{markwardt_non-linear_2009}
Markwardt, C.~B. 2009,
  \href{http://adsabs.harvard.edu/abs/2009ASPC..411..251M}{in Astronomical Data
  Analysis Software and Systems {XVIII}, Vol. 411}, 251

\bibitem[{Masjedi {et~al.}(2008)Masjedi, Hogg, \&
  Blanton}]{masjedi_growth_2008}
Masjedi, M., Hogg, D.~W., \& Blanton, M.~R. 2008,
  \href{http://adsabs.harvard.edu/abs/2008ApJ...679..260M}{The Astrophysical
  Journal, 679, 260}

\bibitem[{Masjedi {et~al.}(2006)Masjedi, Hogg, Cool, Eisenstein, Blanton,
  Zehavi, Berlind, Bell, Schneider, Warren, \& Brinkmann}]{masjedi_very_2006}
Masjedi, M., Hogg, D.~W., Cool, R.~J., {et~al.} 2006,
  \href{http://adsabs.harvard.edu/abs/2006ApJ...644...54M}{The Astrophysical
  Journal, 644, 54}

\bibitem[{Mulchaey(2000)}]{mulchaey_x-ray_2000}
Mulchaey, J.~S. 2000,
  \href{http://adsabs.harvard.edu/abs/2000ARA%26A..38..289M}{Annual Review of
  Astronomy and Astrophysics, 38, 289}

\bibitem[{Mulchaey {et~al.}(1993)Mulchaey, Davis, Mushotzky, \&
  Burstein}]{mulchaey_diffuse_1993}
Mulchaey, J.~S., Davis, D.~S., Mushotzky, R.~F., \& Burstein, D. 1993,
  \href{http://adsabs.harvard.edu/abs/1993ApJ...404L...9M}{The Astrophysical
  Journal Letters, 404, L9}

\bibitem[{Navarro {et~al.}(1996)Navarro, Frenk, \&
  White}]{navarro_structure_1996}
Navarro, J.~F., Frenk, C.~S., \& White, S. D.~M. 1996,
  \href{http://adsabs.harvard.edu/abs/1996ApJ...462..563N}{The Astrophysical
  Journal, 462, 563}

\bibitem[{Navarro {et~al.}(1997)Navarro, Frenk, \&
  White}]{navarro_universal_1997}
---. 1997, \href{http://adsabs.harvard.edu/abs/1997ApJ...490..493N}{The
  Astrophysical Journal, 490, 493}

\bibitem[{Nierenberg {et~al.}(2011)Nierenberg, Auger, Treu, Marshall, \&
  Fassnacht}]{nierenberg_luminous_2011}
Nierenberg, A.~M., Auger, M.~W., Treu, T., Marshall, P.~J., \& Fassnacht, C.~D.
  2011, \href{http://adsabs.harvard.edu/abs/2011ApJ...731...44N}{The
  Astrophysical Journal, 731, 44}

\bibitem[{Reid \& Spergel(2009)}]{reid_constraining_2009}
Reid, B.~A., \& Spergel, D.~N. 2009,
  \href{http://adsabs.harvard.edu/abs/2009ApJ...698..143R}{The Astrophysical
  Journal, 698, 143}

\bibitem[{Sales {et~al.}(2007)Sales, Navarro, Lambas, White, \&
  Croton}]{sales_satellite_2007}
Sales, L.~V., Navarro, J.~F., Lambas, D.~G., White, S. D.~M., \& Croton, D.~J.
  2007, \href{http://adsabs.harvard.edu/abs/2007MNRAS.382.1901S}{Monthly
  Notices of the Royal Astronomical Society, 382, 1901}

\bibitem[{Skelton {et~al.}(2011)Skelton, Bell, \&
  Somerville}]{skelton_modeling_2011}
Skelton, R.~E., Bell, E.~F., \& Somerville, R.~S. 2011, Modeling the red
  sequence: Hierarchical growth yet slow luminosity evolution

\bibitem[{Tal \& van Dokkum(2011)}]{tal_faint_2011}
Tal, T., \& van Dokkum, P.~G. 2011,
  \href{http://adsabs.harvard.edu/abs/2011ApJ...731...89T}{The Astrophysical
  Journal, 731, 89}

\bibitem[{Tal {et~al.}(2009)Tal, van Dokkum, Nelan, \&
  Bezanson}]{tal_frequency_2009}
Tal, T., van Dokkum, P.~G., Nelan, J., \& Bezanson, R. 2009,
  \href{http://adsabs.harvard.edu/abs/2009AJ....138.1417T}{The Astronomical
  Journal, 138, 1417}

\bibitem[{Tal {et~al.}(2012)Tal, Wake, van Dokkum, van~den Bosch, Schneider,
  Brinkmann, \& Weaver}]{tal_mass_2012}
Tal, T., Wake, D.~A., van Dokkum, P.~G., {et~al.} 2012,
  \href{http://adsabs.harvard.edu/abs/2012ApJ...746..138T}{The Astrophysical
  Journal, 746, 138}

\bibitem[{Tojeiro \& Percival(2011)}]{tojeiro_disentangling_2011}
Tojeiro, R., \& Percival, W.~J. 2011,
  \href{http://adsabs.harvard.edu/abs/2011MNRAS.417.1114T}{Monthly Notices of
  the Royal Astronomical Society, 417, 1114}

\bibitem[{Vader \& Sandage(1991)}]{vader_small-scale_1991}
Vader, J.~P., \& Sandage, A. 1991,
  \href{http://adsabs.harvard.edu/abs/1991ApJ...379L...1V}{The Astrophysical
  Journal Letters, 379, L1}

\bibitem[{van~den Bosch {et~al.}(2005)van~den Bosch, Yang, Mo, \&
  Norberg}]{van_den_bosch_abundance_2005}
van~den Bosch, F.~C., Yang, X., Mo, H.~J., \& Norberg, P. 2005,
  \href{http://adsabs.harvard.edu/abs/2005MNRAS.356.1233V}{Monthly Notices of
  the Royal Astronomical Society, 356, 1233}

\bibitem[{van Dokkum {et~al.}(2010)van Dokkum, Whitaker, Brammer, Franx, Kriek,
  Labbé, Marchesini, Quadri, Bezanson, Illingworth, Muzzin, Rudnick, Tal, \&
  Wake}]{van_dokkum_growth_2010}
van Dokkum, P.~G., Whitaker, K.~E., Brammer, G., {et~al.} 2010,
  \href{http://adsabs.harvard.edu/abs/2010ApJ...709.1018V}{The Astrophysical
  Journal, 709, 1018}

\bibitem[{Vikhlinin {et~al.}(2006)Vikhlinin, Kravtsov, Forman, Jones,
  Markevitch, Murray, \& Van~Speybroeck}]{vikhlinin_chandra_2006}
Vikhlinin, A., Kravtsov, A., Forman, W., {et~al.} 2006,
  \href{http://adsabs.harvard.edu/abs/2006ApJ...640..691V}{The Astrophysical
  Journal, 640, 691}

\bibitem[{Wake {et~al.}(2008)Wake, Sheth, Nichol, Baugh, {Bland-Hawthorn},
  Colless, Couch, Croom, de~Propris, Drinkwater, Edge, Loveday, Lam, Pimbblet,
  Roseboom, Ross, Schneider, Shanks, \& Sharp}]{wake_2df-sdss_2008}
Wake, D.~A., Sheth, R.~K., Nichol, R.~C., {et~al.} 2008,
  \href{http://adsabs.harvard.edu/abs/2008MNRAS.387.1045W}{Monthly Notices of
  the Royal Astronomical Society, 387, 1045}

\bibitem[{Watson {et~al.}(2011)Watson, Berlind, {McBride}, Hogg, \&
  Jiang}]{watson_extreme_2011}
Watson, D.~F., Berlind, A.~A., {McBride}, C.~K., Hogg, D.~W., \& Jiang, T.
  2011, The Extreme Small Scales: Do Satellite Galaxies Trace Dark Matter?

\bibitem[{York {et~al.}(2000)York, Adelman, Anderson, Anderson, Annis, Briegel,
  Briggs, Brinkmann, Brunner, \& Burles}]{york_sloan_2000}
York, D.~G., Adelman, J., Anderson, J.~E., {et~al.} 2000,
  \href{http://adsabs.harvard.edu/abs/2000AJ....120.1579Y}{The Astronomical
  Journal, 120, 1579}

\bibitem[{Zehavi {et~al.}(2002)Zehavi, Blanton, Frieman, Weinberg, Mo, Strauss,
  Anderson, Annis, Bahcall, Bernardi, Briggs, Brinkmann, Burles, Carey,
  Castander, Connolly, Csabai, Dalcanton, Dodelson, Doi, Eisenstein, Owen, \&
  Vogeley}]{zehavi_galaxy_2002}
Zehavi, I., Blanton, M.~R., Frieman, J.~A., {et~al.} 2002,
  \href{http://adsabs.harvard.edu/abs/2002ApJ...571..172Z}{The Astrophysical
  Journal, 571, 172}

\bibitem[{Zheng {et~al.}(2009)Zheng, Zehavi, Eisenstein, Weinberg, \&
  Jing}]{zheng_halo_2009}
Zheng, Z., Zehavi, I., Eisenstein, D.~J., Weinberg, D.~H., \& Jing, Y.~P. 2009,
  \href{http://adsabs.harvard.edu/abs/2009ApJ...707..554Z}{The Astrophysical
  Journal, 707, 554}

\end{thebibliography}
\bibliographystyle{yahapj}

\end{document}